\documentclass[12pt]{article}
\usepackage{amsmath,graphicx,subfigure}

\date{\today}

\begin{document}
\author{Roberto Brambilla$^1$,Francesco Grilli$^2$ \\
$^1${\small RSE -- Ricerca sul Sistema Energetico, Milan, Italy} \\
$^2${\small KIT -- Karlsruhe Institute of Technology, Karlsruhe, Germany}
}
\title{Critical state solution of a cable made of curved thin superconducting tapes}

\date{}

\maketitle

\begin{abstract}
In this paper we develop a method based on the critical state for calculating the current and field distributions and the AC losses in a cable made of curved thin superconducting tapes. The method also includes the possibility of considering spatial variation of the critical current density, which may be the result of the manufacturing process. For example, rare-earth based coated conductors are known to have a decrease of the transport properties in near the edges of the tape: this influences the way current and field penetrate in the sample and, consequently, the AC losses. We demonstrate that curved tapes arranged on a cylindrical former behave as an infinite horizontal stack of straight tapes, and we compare the AC losses in a variety of working conditions, both without and with the lateral dependence of the critical current density. This model and subsequent similar approaches can be of interest for various applications of coated conductors, including power cables and conductor-on-round-core (CORC) cables.\end{abstract}


\section{Introduction}
Superconducting tapes are often assembled in cable configurations in order to form conductors with the high current capacity. The performance of such cables strongly depends on the electromagnetic interaction of the tapes and it is therefore important to have simple and fast numerical tools able to evaluate that interaction.
In a recent paper~\cite{Brambilla:APL13} we considered the problem of the critical state for superconducting thin tapes forming a polygonal cable. That model is based on the Biot-Savart law for current distributions with angular periodicity and on its transformation into a Cauchy singular integral equation.
As regards the AC losses, we found that two major factors greatly influence their value: (a) the number of tapes (edges of the polygon) and (b) the lateral gap between the tapes. In this paper, we utilize the same approach for a system of identical curved tapes, which we call for brevity an {\it arc-polygonal cable}. This configuration describes more accurately the curvature of the tapes in conductor-on-round-core (CORC) cables~\cite{VanderLaan:SST11} and in other conductor concepts featuring curved tapes~\cite{Ma:SST06, Allais:USPatent14}.

\section{Definition of the problem}

\begin{figure}[b!]
\begin{center}
\includegraphics[width=8cm]{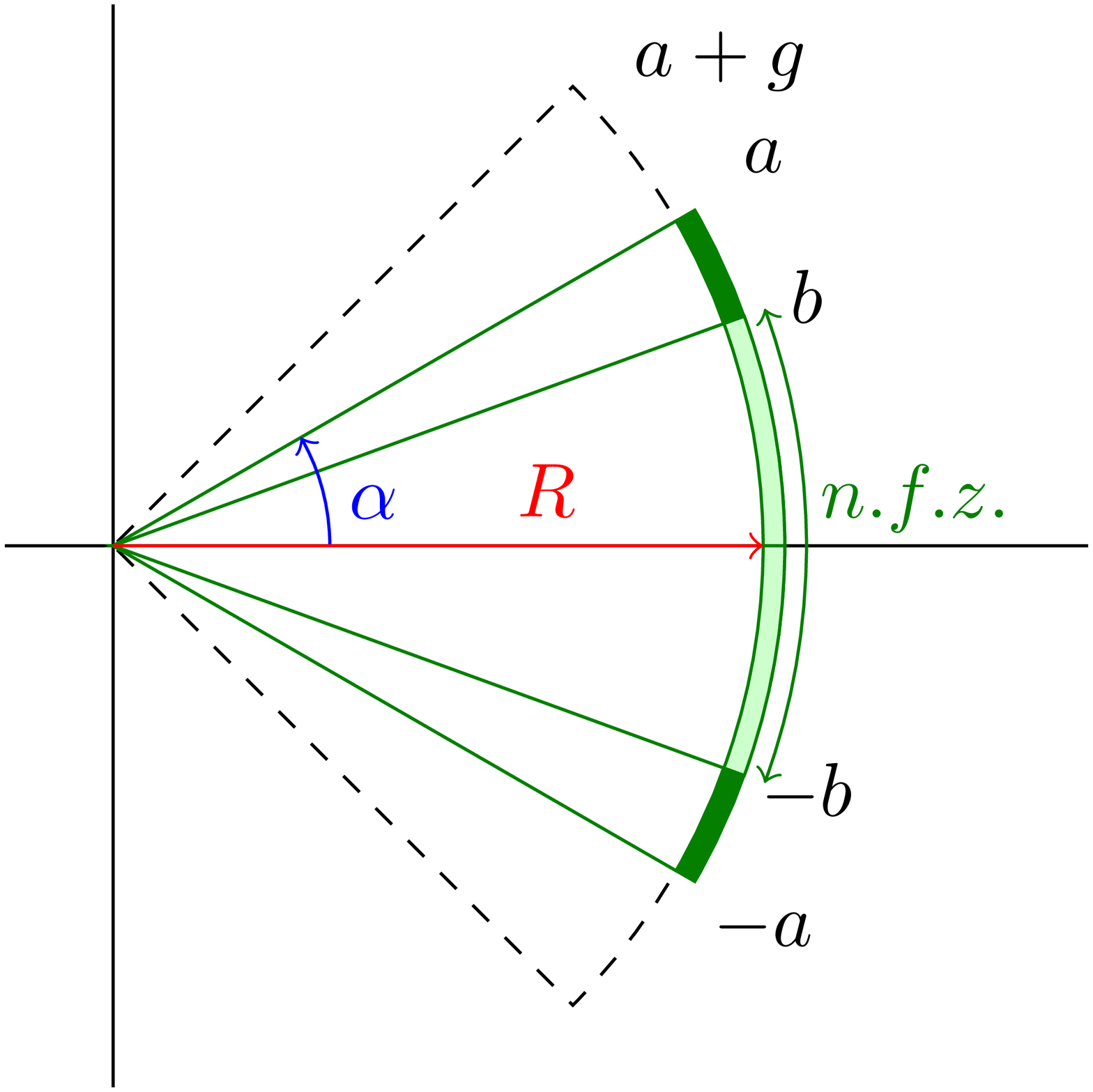}
\end{center}
\caption{Geometry of the arc-polygonal cable (only one tape shown). The region in the interval $(-b,b)$ is the null-field zone ($n.f.z.$), where there is no penetration of the magnetic field.}\label{fig:geometry}
\end{figure}

Let us consider a set of $n$ equidistant arches, each of length $2a$, positioned on a circumference of radius $R$ and separated by a (circular) gap $2g$, as displayed in figure~\ref{fig:geometry}. Each tape carries the same sheet current density $J(\zeta)$, where $\zeta$ is the source point moving along the arc $\Gamma: (-a,a)$ representing the cross-section of the conductor.
Applying the Biot-Savart law for current distributions with angular periodicity, one can express the magnetic field in complex form as~\cite{Brambilla:APL13}
\begin{equation}\label{eq:Biot-Savart}
H(z)=H_y(z)+iH_x(z)=\frac{1}{2\pi}\int_{\Gamma}J(\zeta)\frac{n z^{n-1}}{z^n-\zeta^n}d\Gamma.
\end{equation}
Defining $L=a+g$ the radius of the supporting circle is $R=n L / \pi$ and the half angle of the tape sector is $\alpha=a/R$. One can rewrite~(\ref{eq:Biot-Savart}) as a function of the polar angle posing $\zeta=Re^{i\theta}$
\begin{equation}
H(z)=\frac{n}{2\pi}\int_{-\alpha}^{\alpha}J(\theta)\frac{z^{n-1}}{z^n-{R}^n e^{in\theta}}R d\theta.
\end{equation}
If one considers in particular a field point on the arc, the integral must be calculated as principal value: defining $z=Re^{i\phi}$, the magnetic field is
\begin{align}
\nonumber
H(\phi)
&=\frac{n}{2\pi}{\rm p.v.}\int_{-\alpha}^{\alpha}J(\theta)\frac{R^{n-1}e^{i(n-1)\phi}}{R^ne^{in\phi}-R^ne^{in\theta}}R d\theta \\
&=\frac{n}{2\pi}{\rm p.v.}\int_{-\alpha}^{\alpha}J(\theta)\frac{e^{i(n-1)\phi}}{e^{in\phi}-e^{in\theta}} d\theta.
\end{align}
By using the complex form of the magnetic field, the normal to the arc $H_{\perp}(\phi)={\bf n}(\phi) \cdot {\bf H}(\phi)=\cos \phi H_x(\phi)+\sin \phi H_y(\phi)$ is given by the imaginary part of $e^{i\phi}H({\phi})$, i.e.
\begin{align}
\nonumber
H_{\perp}(\phi)
&=\frac{n}{2\pi}{\rm p.v.}\int_{-\alpha}^{\alpha}J(\theta)\Im \left [ \frac{1}{1-e^{in(\theta-\phi)}}\right ] d\theta \\
&=\frac{n}{2\pi}{\rm p.v.}\int_{-\alpha}^{\alpha}J(\theta)\frac{1}{2}\cot \left [ \frac{1}{2} n (\theta - \phi )\right] d \theta
\end{align}
Introducing the arc coordinates $t=R \theta$  and $s=R \phi$, by simple substitution, since $d\theta=dt/R=\pi dt/nL$, we obtain the final expression
\begin{equation}\label{eq:introHperp}
H_{\perp}(s)
=\frac{1}{2L}{\rm p.v.}\int_{-a}^{a}K(s,t)J(t)dt,
\end{equation}
where we have defined the Green function for the arc-polygonal 
\begin{equation}\label{eq:Kst}
K(s,t)=\frac{1}{2}\cot \frac{\pi (s-t)}{2L}.
\end{equation}
Equation~\eqref{eq:introHperp} tells us that the field $H_{\perp}$ is the  {\it finite} Hilbert transform of the current density $J$.

This expression is independent of $n$, i.e. the perpendicular field on the arc $(-a,a)$ does not depend on the number of arches composing the cable. Assuming a number $n$ approaching infinity, one also has that $R$ approaches infinity, so that the cable becomes an infinite horizontal array (the so-called $X$-array~\cite{Muller:PhysC97a}). Therefore we can conclude that the distribution of the perpendicular field in a cable made of curved tapes does not depend on $n$ and it is equal to that of the perpendicular  field of an $X$-array made of the same tapes and gaps.
The circumstance $n=1$ relates to the case of a single tape with curvature of radius $R$, where $2g$ is the remaining part of the circumference.

The independence of the results on the number tapes $n$ had been already derived, although perhaps little emphasized, by Mawatari~\cite{Mawatari:PRB09}.
He applies a method based on conformal mapping of exponential type that converts the curved tapes into an infinite $X$-array. In the transformed complex plane the magnetic field is then obtained by applying the usual Biot-Savart formula and the results are pulled back to the original cable plane by the inverse transform.  In contrast, in our approach the magnetic field is obtained as an immediate application of the Biot-Savart formula for angular-periodic currents, see equation~\eqref{eq:Biot-Savart}, avoiding the detour of complex transforms, which one can consider as hindered in the Hilbert kernel~\eqref{eq:Kst}. Straightforward variable substitutions in the integral lead directly to the independence of the results on the number of tapes $n$.
In addition, as it will be shown later, the fact that $J_c$ may not be uniform, but depend on the position (as a consequence of irregularities and difficulties of the manufacturing process) can be easily implemented in the current approach.

\section{Constant $J_c$ along the whole section of the tape}

As a first application of formula~\eqref{eq:introHperp} we consider the case of superconducting tapes in the critical state characterized by a constant $J_c$.  Since the arc polygonal case coincides with the $X$-array case in this paragraph, for sake of practicality to the reader, we simply report the formulas that have been found in the past, using different methods~\cite{Muller:PhysC97a, Mawatari:PRB09, Brambilla:APL13}. Suppose that $I_c$ is the critical current for each tape and that the magnetic field  penetrates each tape only in the lateral rims $(-a,-b)$  and $(b,a)$ (figure~\ref{fig:geometry}). According to the critical state assumption, in those bands the current density is constant and assumes the critical value $J_c=I_c/2a$. In the central zone  where the field is null (n.f.z.) the current density assumes a profile $g_1(s)$
\begin{align}\label{eq:gs}
J_1(s)=J_c
\begin{cases}
g_1(s)		&	(0 < \left| s \right| < b) \\
1 	&	(b< \left| s \right| < a).\\
\end{cases}
\end{align}
The unknown function $g_1(s)$ satisfies the integral equation 
\begin{equation}
\int_{-b}^{b}K(s,t)g_1(t)dt=f_0(s),
\end{equation}
with the given term 
\begin{equation}\label{eq:f0}
f_0(s)=Ls_0\left ( \frac{s}{L}, \frac{b}{L}\right )-
           Ls_0\left ( \frac{s}{L}, \frac{a}{L}\right )
\end{equation}
where we have defined the auxiliary function

\begin{equation}\label{eq:s0}
s_0(x,y)=\frac{1}{\pi}\ln\frac
{\sin \left [ \pi(y+x)/2 \right]}
{\sin \left [ \pi(y-x)/2 \right]}.
\end{equation}
In this special case the integral equation can be solved analytically and we obtain 
\begin{align}\label{eq:g1s}
g_1(s)=\frac{2}{\pi}\tan^{-1}\sqrt\frac
{\tan^2(\pi a/2L)-\tan^2(\pi b/2L)}
{\tan^2(\pi b/2L)-\tan^2(\pi s/2L)}
& & (|s| < b).
\end{align}
which is the same result obtained for an $X$-array by M\"uller~\cite{Muller:PhysC97a}.
The normal component of the magnetic field in the penetration band $(b,a)$ is obtained from~\eqref{eq:introHperp} using~\eqref{eq:gs}
\begin{align}\label{eq:p}
H_{\perp}(s)=\frac{J_c}{\pi}\tanh^{-1}\sqrt\frac
{\tan^2(\pi s/2L)-\tan^2(\pi b/2L)}
{\tan^2(\pi a/2L)-\tan^2(\pi b/2L)}
& & (b<s<a).
\end{align}
Integrating~\eqref{eq:gs} on the arc $(-a,a)$ we find the total current transported by the tape $I_t$ as a function of the penetration $b$
\begin{equation}\label{eq:p}
p=\frac{I_t}{I_c}=\frac{2L}{\pi a}\cos^{-1}\frac
{\cos(\pi a/2L)}
{\cos(\pi b/2L)}.
\end{equation}
\begin{figure}[t!]
\begin{center}
\includegraphics[width=8cm]{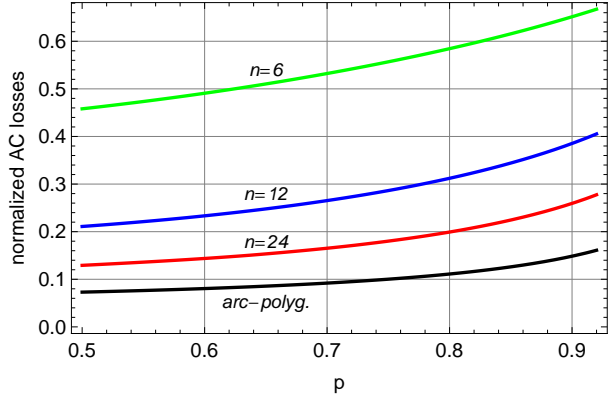}
\end{center}
\caption{AC losses as a function of the normalized current ratio $p$ for a cable with polygonally arranged straight and curved tapes. For straight tapes, the losses depend on the actual number of tapes (in the figure 6, 12, 24); for curved tapes they do not and they are equal to those of an $X$-array.}\label{fig:cfr_polygonal}
\end{figure}
Inverting~\eqref{eq:p} the penetration $b$ can be expressed as a function of the transport current ratio $p$
\begin{equation}\label{eq:b}
b=\frac{2L}{\pi}\cos^{-1}\frac
{\cos(\pi a/2L)}
{\cos(\pi pa/2L)}.
\end{equation}
The AC losses (in each tape of the cable) can be computed resorting to Norris's approach~\cite{Norris:JPDAP70}
\begin{equation}\label{eq:Q_Jc_cst}
Q=\int_0^T dt \int_{-a}^a J_1(s)E(s)ds=8\mu_0J_c\int_b^a (a-s)H_{\perp}(s) ds,
\end{equation}
where $T$ is the period of the AC cycle. This integral appears not to be expressible by means of standard functions but is easily numerically evaluated.

Figure~\ref{fig:cfr_polygonal} shows the AC losses as a function of the transport current for a cable made of $n$ polygonally arranged straight tapes ($n$=6, 12, 24) and of curved tapes. In the latter case the losses are independent of $n$ and are equivalent to those of a tape in an $X$-array configuration. The following parameters were used: tape width=4~mm, gap=0.25 mm. The losses are plotted as a function of the reduced current $p=I_t/I_c$ and are normalized with respect to those of an isolated tape, computed with the well-known formula by Norris~\cite{Norris:JPDAP70}
\begin{equation}
Q=\frac{\mu_0I_c^2}{\pi}\left [
(1+p)\ln(1+p)+(1-p)\ln(1-p)-p^2
\right ].
\end{equation}

\section{Lateral variation of the critical current density}
The assumption of a uniform critical current density along the cross section is mostly motivated by the necessity of simple mathematical handling. However, this is not always the case: for example, HTS coated conductors  have often lower $J_c$ at the edges for reasons linked to the deposition of the superconducting layer in the manufacturing process~\cite{Amemiya:PhysC06b,Grilli:TAS07a,Gomory:TAS13, Solovyov:SST13}. This non-uniformity can importantly influence the tape's performance. In the following we extend our model to take the possibility of non-uniform $J_c$ into account. In particular, we shall consider two cases of non-uniform distribution of $J_c$ for which a reasonable mathematical treatment is still possible.
\subsection{Trapezoidal profile}
\begin{figure}[!t]
\centering
\subfigure[]
{
\label{fig:casetta_plot_0}
\includegraphics[height=6 cm]{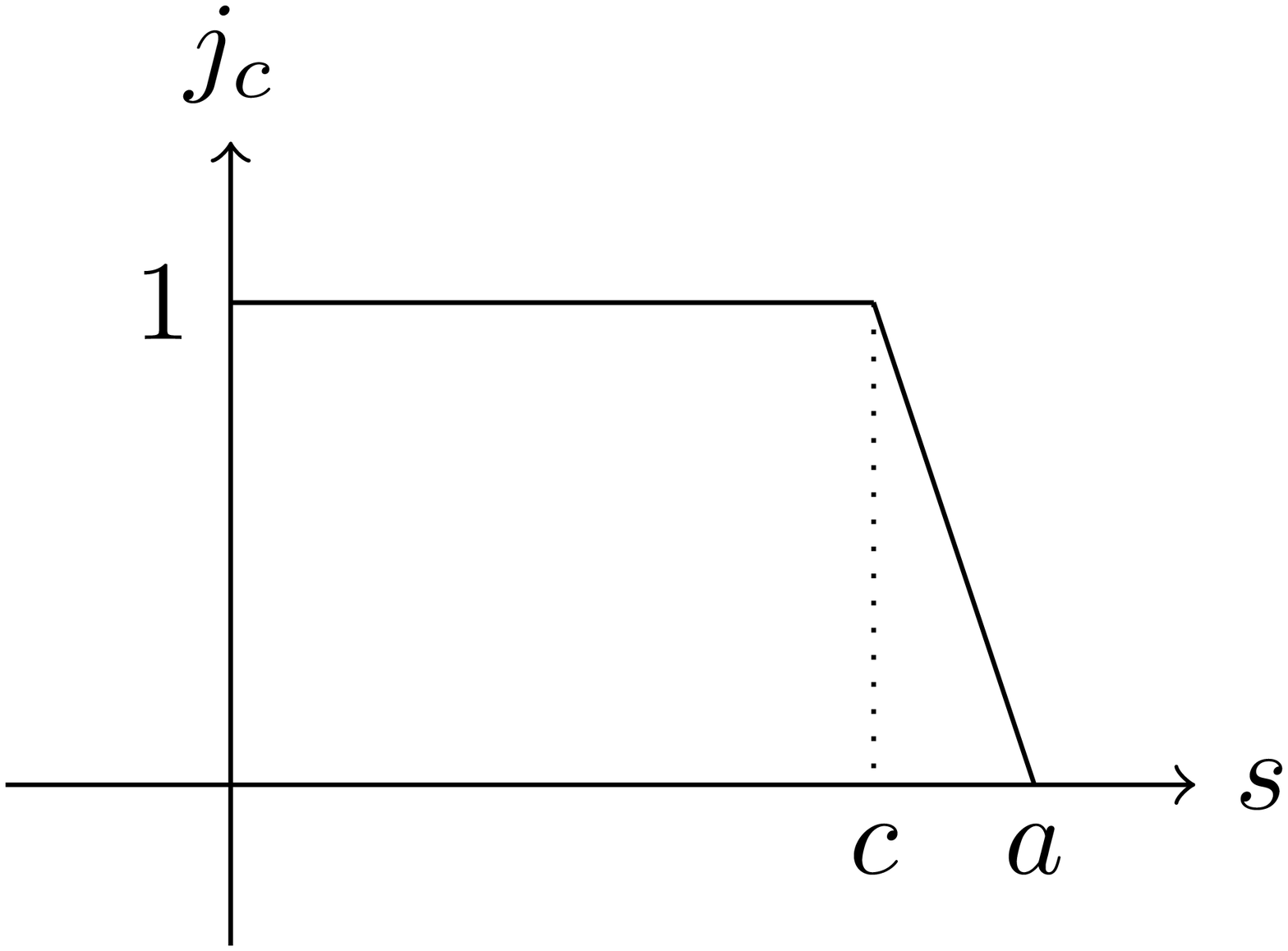}
}
\subfigure[]
{
\label{fig:casetta_plot}
\includegraphics[height=6 cm]{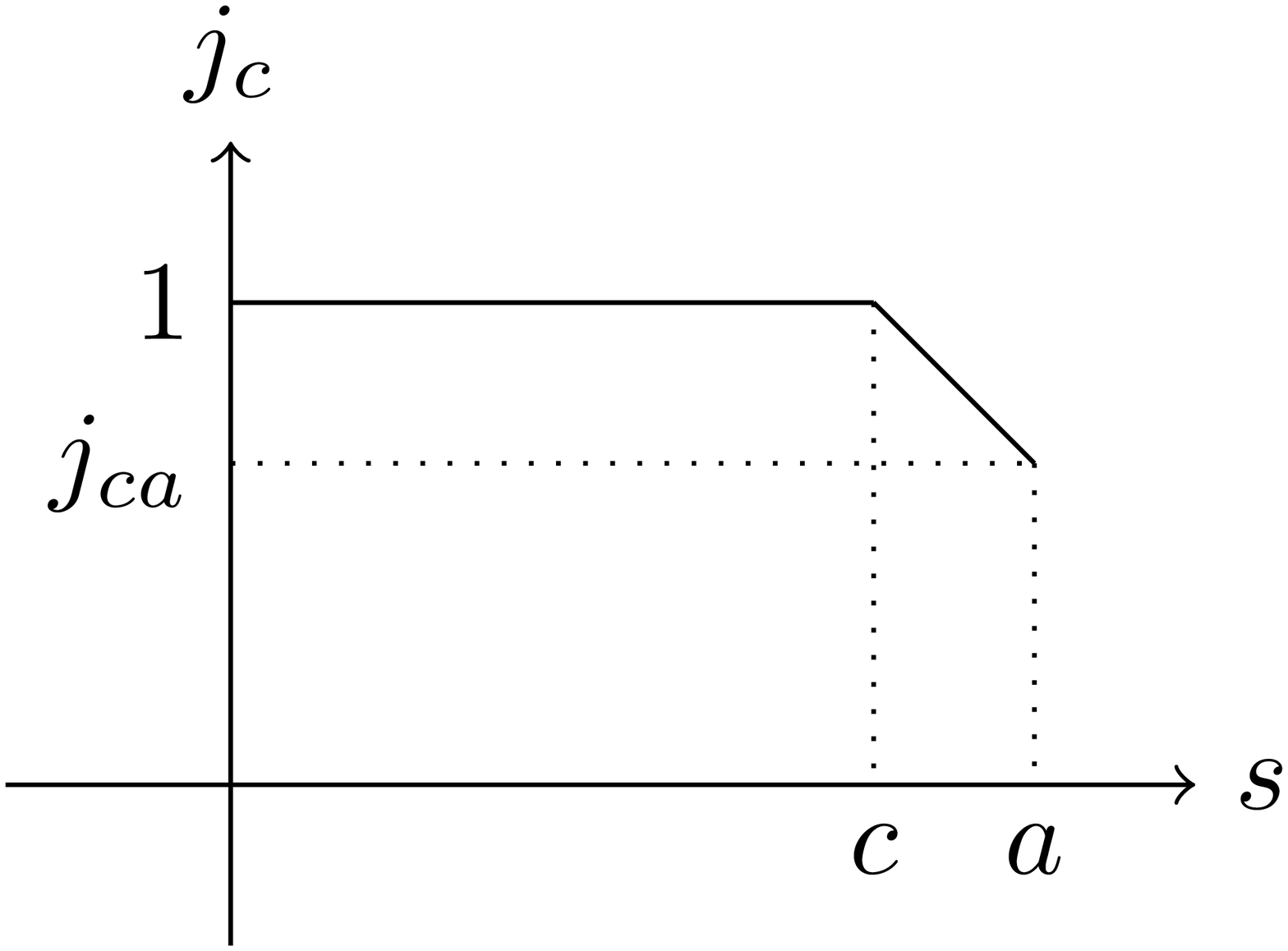}
}
\caption{Trapezoidal variation of the critical current density $j_c(s)$, going to zero (a) or to a finite value $j_{ca}$ (b) at the edges of the tape. For symmetry reason, only the right half of the tape width is shown.}
\end{figure}

First, we shall consider the case of a non-uniform lateral distribution of the critical current density $j_c$, with a linear decrease in the lateral bands $(-a,-c)$ and $(c,a)$
\begin{align}\label{eq:jcTs}
j_{cT}(s)=
\begin{cases}
1 	&	(0 < \left| s \right| < c)\\
T(s)		&	(c < \left| s \right| < a)
\end{cases}
\end{align}
where 
\begin{equation}
T(s)=\frac{|s|-a}{c-a},
\end{equation}
so that the $j_c(s)$ profile has a trapezoidal shape, as shown in figure~\ref{fig:casetta_plot_0}.
The critical current will therefore be $I_c=J_c(a+c)$. Similarly to~\eqref{eq:gs}, the current density is
\begin{align}\label{eq:gs2}
J_2(s)=J_c
\begin{cases}
j_{cT}(s) 	&	(-a < \left| s \right| < -b)\\
g_2(s)		&	(-b< \left| s \right| < b)\\
j_{cT}(s) 	&	(b < \left| s \right| < a)
\end{cases}
\end{align}
where $g_2(s)$ is the unknown function to be determined. The magnetic field is still given by~\eqref{eq:introHperp}
\begin{equation}
{H_ \perp }(s) = \frac{1}{{2L }}{\rm p.v.}\int_{ - a}^a {K(s,t)J_2(t)dt}.
\end{equation}

Two cases need be considered, depending on the null field zone being internal or external to the interval $(-c,c)$. 

Case $b<c$: the magnetic field is null in $(-b,b)$ and one has
\begin{align}
\nonumber
0 &= \int_{ - a}^{ - c} {K(s,t)T(t)} dt + \int_{ - c}^{ - b} {K(s,t)} dt + {\rm p.v.}\int_{ - b}^b {K(s,t)} g_2(t)dt \\
&+ \int_b^c {K(s,t)}dt + \int_c^a {K(s,t) T(t)} dt
\end{align}
or
\begin{equation}\label{eq:pv11}
{\rm p.v.}\int_{ - b}^b {K(s,t)g(t)} dt =   {f_1}(s)
\end{equation}
where the known term is
\begin{equation}\label{eq:f1}
{f_1}(s) = 
Ls_0\left ( \frac{s}{L},\frac{b}{L}\right)+\frac{L^2}{a-c}
\left [
s_1\left ( \frac{s}{L},\frac{a}{L}\right)
-s_1\left ( \frac{s}{L},\frac{c}{L}\right)
\right ],
\end{equation}
where we have defined the auxiliary function
\begin{equation*}
s_1(x,y)=\frac{1}{\pi^2}\Im
\left [  
{\rm Li}_2e^{i\pi(y+x)}-{\rm Li}_2e^{i\pi(y-x)}
\right]
\end{equation*}

(${\rm Li}_2 (z)=\int_z^0 \frac{\ln(1-t)}{t}dt$ is the dilogarithm function and $s_0$ is the same as in~\eqref{eq:s0}).


Case $b>c$: In a completely similar way one will have again~\eqref{eq:pv11}, but with the known term given by
\begin{equation}\label{eq:f2}
{f_2}(s) 
=L\frac{a-b}{a-c}s_0
\left ( \frac{s}{L},\frac{b}{L}\right)
+\frac{L^2}{a-c}
\left [
s_1\left ( \frac{s}{L},\frac{a}{L}\right)
-s_1\left ( \frac{s}{L},\frac{b}{L}\right)
\right ].
\end{equation}
Equation~\eqref{eq:pv11} is singular of Cauchy type (principal value), as it can be seen developing the cotangent in series 
$
K(s,t) = (L/\pi)(s-t)^{-1}+ \sum\limits_{} {{c_k}{{(s - t)}^k}}.
$
A direct solution (of numerical type) is very problematic and it is convenient to extract the singularity from the integral.

Setting $Q(s,t)=(s-t)K(s,t)$ we can write
\begin{align}
\nonumber
& {\rm p.v.}\int_{ - b}^b {K(s,t)} g_2(t)dt 
= {\rm p.v.}\int_{ - b}^b {\frac{{Q(s,t)}}{{s - t}}g_2(t)} dt \\
&= Q(s,s){\rm p.v.}\int_{ - b}^b {\frac{g_2(t)}{{s - t}}} dt 
+ \int_{ - b}^b {\frac{{Q(s,t) - Q(s,s)}}{{s - t}}g_2(t)} dt.
\end{align}
One can immediately note that $Q(s,s)=L/\pi$. Equation~\eqref{eq:pv11} therefore becomes
\begin{equation}\label{eq:to_appendix}
{\rm p.v.}\int_{ - b}^b {\frac{{g_2(t)}}{{t - s}}} dt + \int_{ - b}^b {P(s,t)g_2(t)} dt =   \frac{\pi}{L}f_{1,2}(s)
\end{equation}
where one has defined the non-singular kernel 
\begin{equation*}
P(s,t)=\frac{(\pi/L)Q(s,t)-1}{s-t}.
\end{equation*}
In fact, in virtue of the development of the cotangent, one has $P(s,s)=0$. Equation~\eqref{eq:to_appendix} can now be transformed into a non-singular equivalent one (see appendix for details), more specifically into a Fredholm equation of the second type, which can be easily solved with the usual numerical techniques dedicated to that purpose.
\begin{align}\label{eq:final}
& g_2(s) + \int_{ - b}^b {{K_0}(s,t)g_2(t)} dt = {f_{0,i}}(s) \\
\nonumber
& {K_0}(s,t) = \frac{{\sqrt {{b^2} - {s^2}} }}{{{\pi ^2}}}{\rm p.v.}\int_{ - b}^b {\frac{{P(\sigma ,t)}}{{\sqrt {{b^2} - {\sigma ^2}} }}\frac{{d\sigma }}{{\sigma  - s}}} \\
\nonumber
& {f_{0,i}}(s) = \frac{{\sqrt {{b^2} - {s^2}} }}{{{\pi ^2}}}\frac{\pi}{{L}}{\rm p.v.}\int_{ - b}^b {\frac{{{f_i}(\sigma )}}{{\sqrt {{b^2} - {\sigma ^2}} }}\frac{{d\sigma }}{{\sigma  - s}}} 
\end{align}
Once~\eqref{eq:final} is solved, the distribution of current density along the whole width of the tape is known. One can therefore calculate the total current and the magnetic field.
The AC losses can be obtained by adapting the Norris method to the current case, i.e.
\begin{align*}
Q_2 &=8 \int_0^T {dt} \int_{ - a}^a J_2(s)E(s)ds = 8\int_0^{T/4} {dt} \int_0^a J_2(s)E(s)ds \\ 
&= 8\mu_0\int_0^{T/4}  {dt} \int_b^ads J_2(s) \int_b^s \dot{H}_{\perp}(\sigma,t)d\sigma
= 8\mu_0 \int_b^a ds J_2(s) \int_b^s H_{\perp}(\sigma,T/4)d\sigma.
\end{align*}
Finally, inverting the order of integration, one obtains
\begin{equation}\label{eq:Q2}
Q_2 = 8{\mu _0}\int_b^a d\sigma H_{\perp}(\sigma,T/4)\int_{\sigma}^a J_2(s)ds.
\end{equation}
One can easily verify that in the case of constant $J_c$ in $(b,a)$ one obtains again~\eqref{eq:Q_Jc_cst}.

\subsection{Trapezoidal profile with base}
Let us finally consider the more general case where the distribution of the critical current density $j_c(s)$ does not drop to zero at the edges, but to a finite value $j_{ca}$, as displayed in figure~\ref{fig:casetta_plot}. 
We can treat this case as a weighted superposition of the two previous two cases -- the sum of a constant distribution and a trapezoidal distribution going to zero
\begin{equation}
j_c(s)=j_{ca}+(1-j_{ca})j_{cT}(s),
\end{equation}
with $j_{cT}(s)$ given by~\eqref{eq:jcTs}.
This superposition is possible because the kernel of the respective integral equations is the same~\eqref{eq:Kst}. For this reason the related integral equation will be 
\begin{equation}
\int_{-b}^{b}K(s,t)g(t)dt=j_{ca}f_0(s)+(1-j_{ca})f_{1,2}(s),
\end{equation}
where $f_0(s)$ is given by~\eqref{eq:f0} and $f_{1,2}(s)$ by~\eqref{eq:f1}  and~\eqref{eq:f2}.
By the linearity of this equation, for any chosen value of the penetration $b$, the solution will be given by the same superposition of the solutions of the previous cases
\begin{equation}
g_3(s)=j_{ca}g_1(s)+(1-j_{ca})g_2(s)
\end{equation}
where $g_1(s)$ is given by~\eqref{eq:g1s} and $g_2(s)$ is the solution of~\eqref{eq:final}. The current density will be given by
\begin{align}\label{eq:J3s}
J_3(s)=J_c
\begin{cases}
j_{c}(s) 	&	(-a < \left| s \right| < -b)\\
g_3(s)		&	(-b< \left| s \right| < b)\\
j_{c}(s) 	&	(b < \left| s \right| < a)
\end{cases}
\end{align}
where $J_c$ is a multiplicative constant determined by the request of the actual value of the critical current of the tape $I_c=J_c\int_{-a}^aj_c(s)ds$. 
The transported current  is $I_t=\int_{-a}^aJ_3(s)ds$ and finally we obtain the current ratio $p=I_t/I_c$ as a function of $b$. 
Once the current density is known, the normal magnetic field is given by~\eqref{eq:introHperp} and the AC losses by an integral as in~\eqref{eq:Q2}.

\begin{figure}[!t]
\centering
\includegraphics[width=8 cm]{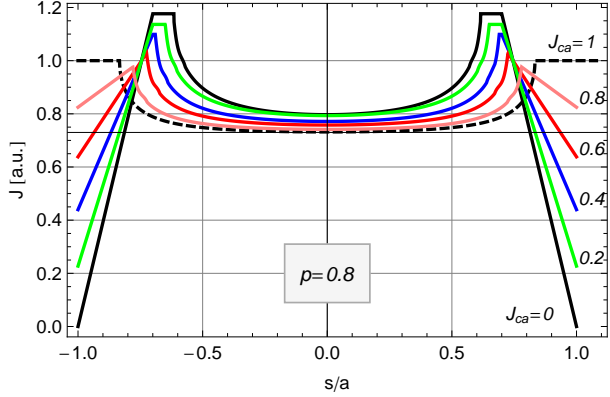}
\caption{\label{fig:profiles}
Comparison of the current density profiles of tapes characterized by different values of $J_{ca}$. The transport and critical currents are 80 and 100~A in all cases ($p=0.8$).}
\end{figure}

\begin{figure}[!t]
\centering

\includegraphics[width=8 cm]{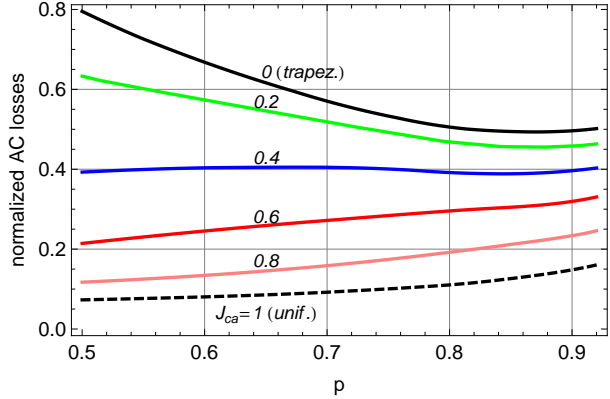}
\caption{\label{fig:losses}Comparison of the transport AC losses of tapes characterized by different values of $J_{ca}$. The losses are normalized with respect to those of a Norris-like strip carrying the same current ratio $p$.}
\end{figure}

As an application, in figure~\ref{fig:profiles} we compare the current density profiles of tapes with the same geometrical properties ($a$=2 mm, $c$=1.4 mm and $g$=0.25 mm), the same critical  and operational current (100 and 80 A, respectively), but different trapezoidal distributions of the critical current density: $j_{ca}$=0, 0.2,0.4, 0.6, 1, the latter case corresponding to a uniform current distribution. It can be clearly seen that the effect of having a non-uniform distribution is to move the region with maximum current toward the center. Another consequence of the non-uniform distribution of $j_c$ is the fact that the losses are higher. This is shown in figure~\ref{fig:losses}, where the losses are plotted as a function of the current ratio $p$. In all cases, the AC losses are normalized to those of a Norris strip characterized by the same given $I_c$ and current ratio $p$. The more pronounced the trapezoidal distribution is (i.e. the lower $j_{ca}$), the higher the losses. Tapes with $j_{ca}$=0.4 have about three times the losses of tapes characterized by a uniform current density distribution.

\section{Conclusion}
In this paper we carried out the critical state solution of curved thin tapes arranged around a cylindrical former. With our method, which avoid complex conformal mapping techniques, we immediately derived the fact that the electromagnetic behavior of this configuration is the same as that of an infinite $X$-array of straight tapes. As such, it does not depend on the number of tapes, but only on the tape width and on the gap between the tapes. The losses of a cable composed of curved tapes are much lower than those of a polygonal configuration, and can be seen as the lowest limit.
We subsequently extended our model to include spatial variations of the critical current density, a situation that for example occurs often in HTS coated conductors. We calculated the current profiles for several spatial distributions of the critical current, and evaluated the impact of such non-uniformity on the AC losses of the cable.
Besides the specific cases considered in this paper, this modeling approach can be extended to other situations of practical interest, characterized by the presence of periodic assemblies of current-carrying tapes.

\newpage
\appendix
\section{Appendix}
The Cauchy type singular equation
\begin{equation}\label{eq:app1}
{\rm p.v} \int_{-b}^{b}\frac{\varphi(t)}{x-t}dt+ \int_{-b}^{b}K(x,t)\varphi(t)dt=f(x)
\end{equation}
where the kernel $K(x,y)$  is not singular, can be rewritten as
\begin{equation}\label{eq:Fx}
{\rm p.v} \int_{-b}^{b}\frac{\varphi(t)}{x-t}dt=F(x)
\end{equation}
where we posed
\begin{equation}\label{eq:Ff}
F(x)=f(x)- \int_{-b}^{b}K(x,\tau)\varphi(\tau)d\tau.
\end{equation}
The solution of~\eqref{eq:Fx} bounded at both extremes is
\begin{equation}
\varphi(x)=\frac{1}{\pi^2}\sqrt{}b^2-x^2 \int_{-b}^{b}\frac{F(t)}{\sqrt{b^2-t^2}}\frac{dt}{t-x}
\end{equation}
with the supplementary constraint that
\begin{equation}
 \int_{-b}^{b}\frac{F(t)}{\sqrt{b^2-t^2}}dt=0.
\end{equation}
By~\eqref{eq:Ff} this request can be satisfied, for instance, if the given term $f(x)$  is an odd function and if the kernel  $K(x,y)$ transforms even functions in odd functions. If this is the case, the solution $\varphi(x)$ turns out to be an even function. Applying this solution to~\eqref{eq:Fx} we obtain
\begin{align}\label{eq:varphi}
\nonumber
\varphi(x)&=\frac{\sqrt{b^2-x^2}}{\pi^2}{\rm p.v.}\int_{-b}^{b}
\frac{f(t)}{\sqrt{b^2-t^2}(t-x)}dt \\
&-\frac{\sqrt{b^2-x^2}}{\pi^2}{\rm p.v.}\int_{-b}^{b}
\frac{1}{\sqrt{b^2-t^2}(t-x)}dt
\int_{-b}^{b}K(t,\tau)\varphi(\tau)d\tau
\end{align}
Changing the integration order in the second term we have
\begin{align*}
& {\rm p.v.}\int_{-b}^{b}
\frac{1}{\sqrt{b^2-t^2}(t-x)}dt
\int_{-b}^{b}K(t,\tau)\varphi(\tau)d\tau \\
&=\int_{-b}^{b}\varphi(\tau)d\tau
\left (
{\rm p.v.} \int_{-b}^{b}\frac{K(t,\tau)}{\sqrt{b^2-t^2}(t-x)}dt
\right )
\end{align*}
so that, defining
\begin{align}
&
f_0(x)=\frac{\sqrt{b^2-x^2}}{\pi^2}{\rm p.v.}\int_{-b}^{b}\frac{f(\tau)}{\sqrt{b^2-\tau^2}(\tau-x)}d\tau \\
& 
\label{eq:K0}
K_0(x,t)=\frac{\sqrt{b^2-x^2}}{\pi^2}{\rm p.v.}\int_{-b}^{b}\frac{K(\tau,t)}{\sqrt{b^2-\tau^2}(\tau-x)}d\tau,
\end{align}
equation~\eqref{eq:varphi} becomes
\begin{equation}
\varphi(x)+\int_{-b}^b K_0(x,t)\varphi(t)dt=f_0(x).
\end{equation}
The singular integral equation~\eqref{eq:app1} -- with the restrictions above -- has the same solution $\varphi(x)$ as the non-singular integral equation~\eqref{eq:K0}. The main advantage of this transformation is that the calculation of the principal value present in the integral equation is shifted to the calculation of a new given term and of a new kernel. A second advantage is that~\eqref{eq:K0} is a Fredholm integral equation of second type, a well posed integral equation for which many solving numerical routines exist (i.e. Nystr\"om methods).

\section*{Acknowledgments}
This work was supported by the Research Fund for the Italian Electrical System under the Contract Agreement between RSE and the Ministry of Economic Development (RB) and by the Helmholtz Association (FG, grant VH-NG-617). 



\end{document}